\documentclass[prb, longbibliography, footinbib,floatfix,twocolumn]{revtex4-1} 
\pdfoutput=1
\usepackage{amsmath} 
\usepackage{geometry}          
\usepackage{graphicx}
\usepackage{multirow} \usepackage{rotating}

\usepackage{amssymb}
\usepackage{natbib}

\graphicspath{{./OerstEPSfigs/}{/Users/chholbrow/Pictures/}{./figuresdir3/}}

\usepackage[OT2,OT1]{fontenc}
\newcommand\cyr{%
\renewcommand\rmdefault{wncyr}%
\renewcommand\sfdefault{wncyss}%
\renewcommand\encodingdefault{OT2}%
\normalfont
\selectfont}
\DeclareTextFontCommand{\textcyr}{\cyr}

\newcommand{\dg}{\mbox{$^{\circ}$}}
\newcommand{\ihat}{\mbox{\,$\hat{\imath}$}}
\newcommand{\jhat}{\mbox{\,$\hat{\jmath}$}}
\newcommand{\khat}{\mbox{\,$\hat{ k}$}}

\newcommand{\pptfig}[2]{\begin{figure}[htbp]  
    \centering
    \includegraphics[width=0.47\textwidth]{#1} 
    \caption{#2}
    \label{#1}
\end{figure}}

\begin{document}
\title{Build Your Own Analemma}
\author{Charles H. Holbrow}
\affiliation{Colgate University \& MIT}
\date{\today}                                           

\begin{abstract}
Earth's analemma is the lopsided figure eight marked out over a year by the position of the Sun in the sky observed at the same clock time each day.  It shows how solar time deviates from clock time.  The analemma's shape results from the tilt of Earth's axis of rotation relative to the plane of its orbit around the Sun and from the elliptical shape of that orbit.  This tutorial paper uses vector analysis of the Earth-Sun geometry and a numerically generated quantitative description of Earth's motion around the Sun to construct Earth's analemma.  To visualize the geometry and the motion that give rise to the analemma is challenging, but this construction is a project within the capability of physics students who have had basic undergraduate mechanics.
\end{abstract}

\maketitle

\section{Introduction}
\pptfig{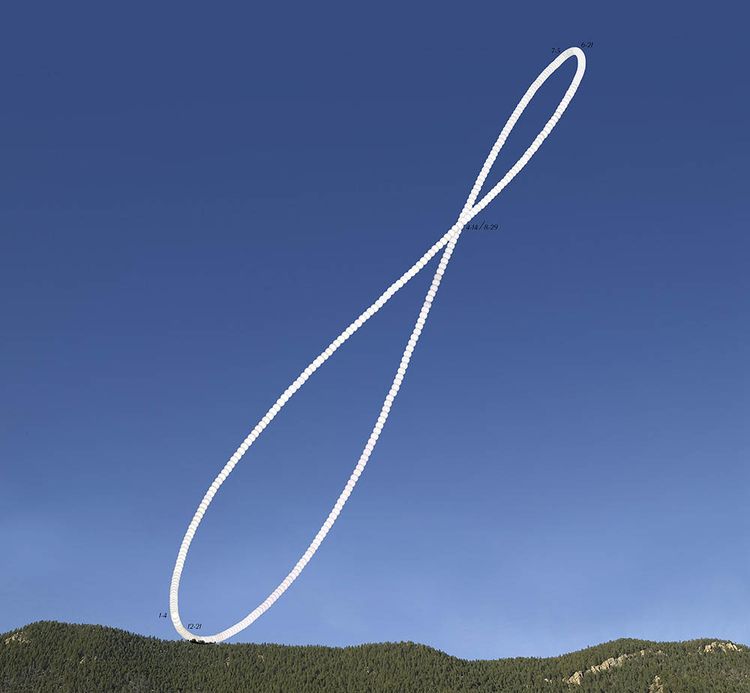}{Earth's analemma as a composite of photographs of the Sun taken at the same location and time every day for a year.} Gary Cooper's shootout in the movie ``High Noon'' almost surely was not at high noon.  He was using clock time as established by the railroads.  Even if he happened to be at the longitude of his time zone corresponding to solar time, there are only four days in the year when the Sun crosses the meridian -- its daily highest point in the sky --  at 12:00:00 clock time.  

During a year solar noon varies from 16~min 33~s earlier to 14~min 6~s later than clock time.   
This variation is apparent in Fig.~\ref{ana_CO_2004} which is a composite of pictures taken in Cascade, Colorado at the same clock time each day  over the course of a year.\cite{Rych:2005}  The curve outlined by these 365 images is Earth's analemma, and the purpose of this paper is to show how to construct it from basic principles of geometry and physics.

The collection of images of the Sun in Fig.~\ref{ana_CO_2004} is low in the western sky because the pictures were taken in the afternoon. The belt of the figure eight, i.e., the points slightly below where the lines of the eight cross, moves along an arc that is the Earth's equator projected onto the sky, i.e., the celestial equator. You can imagine the figure eight of the analemma rising somewhat tilted in the east -- upper side first; then the figure eight, long axis perpendicular to the celestial equator, moves across the sky and sets in the west -- upper side last. At noon the long axis of the figure eight lies along the meridian, half way between the eastern and western horizons.  

The analemma is a record of the Sun's position in the sky as observed from the same point on Earth at time intervals of exactly one day.  The analemma displays the variation of the angle between the Sun's path in the sky and the celestial equator. Angle measured perpendicular from the celestial equator is called declination. The seasonal variation of the Sun's declination from $-23.44\dg$ to $+23.44\dg$ and back again gives the analemma its long dimension.  A small periodic variation in the time of day when the Sun crosses the meridian gives the analemma its width of a few degrees.\footnote{Astronomers call this small variation the ``equation of time,'' using the word ``equation'' in an obsolete sense meaning `` correction.''  The analemma provides the correction to convert sundial time to clock time and vice versa.}

This paper shows how the relative motion of Sun and Earth results in the two angular variations that  produce Earth's analemma. The paper provides an analytic explanation in terms of two useful vectors, one that describes the motion of the actual Sun, and the other that describes a fictitious average or ``mean Sun" that corresponds to clock time. These vectors are defined in section~\ref{vectors}.

Section~\ref{tilt} uses diagrams to show how geometry results in the small periodic variation in the daily time at which the Sun crosses a meridian. These diagrams show how tilt produces an analemma even if the orbit is a perfect circle; they also show how to extract useful information from the two vectors.  The vectors are then used to produce the analemma for the special case of a  circular orbit. The result is a symmetric figure-eight analemma with its mid-point exactly on the celestial equator. 

Section~\ref{ellipse} extends the vector approach to generate the analemma arising from an elliptical orbit.  This section uses a numerical calculation of Earth's orbit and shows how its ellipticity gives the analemma the different sized loops and slightly lopsided shape you see in Fig~\ref{ana_CO_2004}.  The excellent agreement of the calculated analemma with the observed one is confirmed by laying one over the other.

A higher level of vector analysis is exhibited in section~\ref{Vectoria}  by using vector algebra to construct the analemma.  The outcome is elegant and compact. 

In sections \ref{projects} and \ref{conclusions} are a few fun facts for fostering facultative physics fora, some suggestions for student projects that could use extensions of the ideas and techniques developed here, and some general conclusions. 

\nocite{revtex41control}\nocite{apsrev41control}

\section{Two vectors for constructing an analemma \label{vectors}}

It is the variation between clock time and Sun time that gives rise to the analemma. Consequently, to construct the analemma it is useful to define two vectors. One,  $\vec{r}_{\rm S}(t)$, corresponding to the Sun's observed motion; the other, $\vec{r}_{\rm m}(t)$, corresponding to the motion of the fictional mean Sun and thus to the passage of clock time $t$.

The vector $\vec{ r}_{\rm S}(t)$ points at the real Sun as it makes its passage across the sky.  The vector varies seasonally both in position and rate as it tracks the apparent geocentric motion of the Sun. 

The vector $\vec{r}_{\rm m}(t)$ points at the mean Sun.  Because the mean Sun defines the rate at which clocks run, $\vec{r}_{\rm m}(t)$ is in effect a clock vector. Thus, if $\vec{r}_{\rm m}(t)$ is set up on the equator at the longitude used to define a time zone, it will point at the mean Sun's location directly overhead at exactly 12:00:00 noon in that time zone. For someone living in the U.S. Mountain Time Zone (UTC-07) that longitude is 105\;\dg\!W.  Set on the equator at that longitude, the vector $\vec{r}_{\rm m}(t)$ will point directly overhead at the fictitious mean Sun as it crosses the meridian of people observing from 105\;\dg\!W longitude --- for example, from downtown Denver, CO. 

  As you will see, the analemma's shape is implicit  in the difference between the two vectors:
\begin{equation}
\vec{\cal{A}}(t) = \vec{r}_{\rm S}(t) - \vec{r}_{\rm m}(t). \label{eq:A}
\end{equation}

 Figure~\ref{ana_RmRS2b} shows the positions of the Sun and the mean Sun around a planet at successive times. (To make the tilt more evident, it has been drawn as $45\,\dg.$) The dots are locations of the tips of the vectors $\vec{r}_{\rm S}(t)$ and $\vec{r}_{\rm m}(t)$ at 14-day intervals. Thus, the figure maps two spatial and one time dimensions into three spatial dimensions. The connected dots correspond to the angular position (two spatial dimensions) of the mean Sun exactly over a particular meridian --- the meridian of the observer --- day after day, and the arc length between dots of the mean Sun corresponds to an exactly integer number of days (the time dimension). The dots of the actual Sun are spaced by the same time interval as those of the mean Sun, but, as you will see, except in four instances the actual Sun can not cross the observer's meridian at the same time as the mean Sun. The dots of the real Sun move back and forth relative to the dots of the mean Sun.

\pptfig{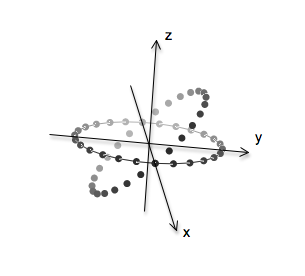}{Intersecting circular paths traced out by the tip of the mean Sun (clock) vector --- shown by connected dots --- and the tip of the Sun vector for a circular orbit.  The points of intersection of the two circles correspond to Earth-Sun equinoxes.  The axes are oriented so the $x$-axis lies on the line of intersection (the line of nodes) of the two circles; the $z$-axis is perpendicular to the plane of the mean Sun and, therefore, parallel to the planet's axis of rotation. For visual impact the plane of $\vec{r}_{\rm S}$ is here tilted $45\dg$ to the equatorial plane swept out by $\vec{r}_{\rm m}$; the dots are 14 days apart if the orbital period is a year. } 
To extract the analemma from $\vec{r}_{\rm S}(t)$ and $\vec{r}_{\rm m}(t)$  you need to represent them in a coordinate system.  It is customary to use the projection of Earth's system of spherical coordinates (latitude and longitude) onto the sky.  For example, the circular arc  formed by projecting Earth's equator onto the sky defines the celestial equator. 

Other projected lines of Earth's latitude define corresponding lines of celestial latitude, i.e., declination.  Lines of celestial longitude are defined similarly by projecting Earth's lines of longitude onto the sky. 

 Because of Earth's rotation, the lines of celestial longitude rise in the east, sweep overhead, and set in the west.  By convention the line of zero celestial longitude lies in the plane defined by the Earth's axis and the line of nodes.  

To represent  $\vec{r}_{\rm S}(t)$ and $\vec{r}_{\rm m}(t)$ it is convenient to use the celestial spherical polar coordinates in a Cartesian reference frame with its origin at Earth, its $z$-axis along Earth's axis of rotation, and its $x$- and $y$-axes in the equatorial plane.  Referring to Fig.~\ref{ana_RmRS2b}, you see that the $x$- and $y$-axes lie in the plane of the circle of connected dots traced out by $\vec{r}_{\rm m}(t)$. 

By convention the $x$-axis is set to lie along the line of intersection (called the line of nodes) of the two circles and to point toward the point where the rising $\vec{r}_{\rm S}(t)$ of the tilted circle crosses the horizontal circle.  In the geometry of Earth and Sun this crossing occurs around March 20 --- the spring equinox of the Northern Hemisphere.

In this frame of reference $\vec{r}_{\rm m}(t)$ tracing out a circle in the $x$-$y$ plane (Fig.~\ref{ana_RmRS2b}) can be written
\begin{equation}
\vec{r}_{\rm m}(t) = \cos \omega t \ihat + \sin \omega t \jhat, \label{eq:rm}
\end{equation}
where $\ihat$ and $\jhat$ are unit vectors along the $x$- and $y$-axes respectively, $t$ is time, and $\omega$ is the angular velocity of Earth on its axis.  The time coordinate is measured from $t=0$ at the March equinox.

The time dependence of $\vec{r}_{\rm m}(t)$ is made to be such that after exactly one day, it again points directly overhead at the mean Sun as it crosses the observer's meridian.  This time interval between two successive occurrences of the mean Sun crossing any chosen meridian is defined to be 24~h, the internationally agreed upon clock length of an Earth day.  Clocks are built to tick off exactly 24~h from one meridian crossing of the mean Sun to the next.

 It is important to understand that because Earth is traveling around the Sun, $\vec{r}_{\rm m}(t)$ rotates through more than $2\pi$~radians in a day.  This extra rotation occurs because  even if Earth were not rotating on its axis, over the course of a year (365.25~d) the Sun and $\vec{r}_{\rm m}(t)$ would still make a rotation of $2\pi$~radians around Earth.  

To see the origin of this extra rotation, imagine Earth did not rotate on its axis at all.  Then if on March 20 (for example) $\vec{r}_{\rm m}(t)$  points at the Sun, six months later it will point directly away from the Sun, and after a year, it will once again point at the Sun.  Because of this extra rotation, in a year of 365.25~d  an observer on Earth sees the Sun circle the Earth 366.25 times. Because of this extra rotation, the celestial longitude of the mean Sun observed from Earth advances by $\omega_0 = 2 \pi/365.25 = 0.01720$~rad each day.
 
To take into account this extra rotation, $\vec{r}_{\rm m}(t)$ must rotate through an angle of $2\pi + 2 \pi/ 365.25$~radian each day.  Then after $n$ days of rotation $\vec{r}_{\rm m}(t)$ will have rotated through an angle of $n 2 \pi + n \omega_0$. If the tip of $\vec{r}_{\rm m}(t)$ made a dot every 14 days, the dots would form a circle like the connected dots in Fig.~\ref{ana_RmRS2b}.

If Earth's orbit around the Sun were circular and if Earth's axis were not tilted relative to the plane of the orbit, the  Sun would move in the sky exactly like the mean Sun; $\vec{r}_{\rm S}(t)$ would trace out a circle exactly like that traced out by $\vec{r}_{\rm m}(t)$. To see that this is so, relate $\vec{r}_{\rm S}(t)$ to $\vec{r}_{\rm m}(t)$ by expressing the time $t$ as the number of days $n$. Then for no tilt and a circular orbit, the Sun vector is
\begin{equation} 
\vec{r}_{\rm S}(n) = \cos n \omega_0  \ihat + \sin n \omega_0  \jhat. \label{eq:S}
\end{equation}
   The clock vector rotates faster, but, as discussed above,  every 24~h it points at the mean Sun crossing the same meridian that it crossed a day earlier.  It is the same meridian, but that meridian is aligned with an arc of celestial longitude $\omega_0 = .01720$~rad ahead of the one it aligned with the day before.  As a result,
\begin{eqnarray}
\vec{r}_{\rm m}(n) & = & \cos (n2 \pi + n\omega_0) \ihat \nonumber \\
 & \ \ \ \ +& \sin(n2 \pi + n \omega_0) \jhat. 
\end{eqnarray}
Because $\cos (n2 \pi + n \alpha) = \cos(n \alpha)$ (and similarly for the sine), it follows that every 24~h $\vec{r}_{\rm m}(n)  = \vec{r}_{\rm S}(n)$, i.e.,
\begin{eqnarray}
\vec{r}_{\rm m}(n) & = & \cos n \omega_0 \ihat + \sin n \omega_0 \jhat \label{eq:rm} \\
& = &\vec{r}_{\rm S}(n). \nonumber
\end{eqnarray}
Thus, for an untilted circular orbit the analemma would be a single image of the Sun always at the same place in the sky at the same time each day. 

\section{ Effect of Axis Tilt on the Analemma \label{tilt}}
 Because Earth's axis of rotation is tilted 23.44\;\dg relative to the plane of its orbit around the Sun, the equality of Eq.~\ref{eq:rm} does not hold for the real Sun. $\vec{r}_{\rm S}(t)$ traces out a path at an angle to the path traced out by $\vec{r}_{\rm m}(t)$.  The tips of the two vectors trace out paths similar to those shown in Fig.~\ref{ana_RmRS2b}.

The analemma is shaped by the tilt of Earth's axis of rotation and by the ellipticity and orientation of its orbit.\footnote{Astronomers call the tilt angle the ``obliquity'' of Earth's rotation axis.}   The major effect comes from tilt. This is apparent when you construct an analemma for an orbit that is a circle.  A tilted, circular orbit is also easier to deal with than an elliptical orbit, so it's a good procedure to first try out $\vec{r}_{\rm S}(t)$  and $\vec{r}_{\rm m}(t)$ in this simpler context and then modify them to take into account the actual orientation and elliptical shape of Earth's orbit. 

Because of the tilt of Earth's axis, the Sun's angle relative to the celestial equator (its declination) changes over the course of a year.  Thus, for Earth, the points of $\vec{r}_{\rm S}(t)$ on the tilted circle trace out the seasonal movement of the Sun as it goes from a declination of $\delta = 0\dg$ at the March equinox to $\delta = 23.44$\dg (the Northern Hemisphere's summer solstice) to 0\;\dg (autumnal equinox) to $-23.44$\;\dg (winter solstice) and so on. Figure~\ref{ana_RmRS2b} illustrates the variation in declination but with the tilt set to $45$\;\dg to make the effect more apparent. 


It is because its path is at an angle to the equatorial plane that the real Sun arrives over a given meridian at a different clock time each day.  That is to say, when the mean Sun crosses a meridian, the real Sun will be over a slightly different meridian. Because Earth rotates at $15\,\dg$ per hour, the angular difference between the longitudes of these two meridians corresponds to a time difference.  For example, on a day when it is exactly clock noon at your meridian and the real Sun is overhead a meridian $1\,\dg$ to the east, it will cross your meridian  4~min later at 12:04 pm.  A difference of  $1\,\dg$ in longitude corresponds to 4~min of time, and 1~arc-min corresponds to  4~s of time.  

\pptfig{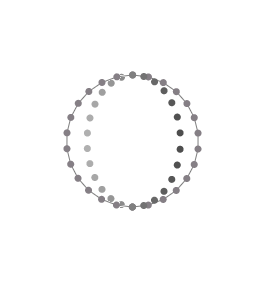}{This is the appearance of the two intersecting circles shown in Fig.~\protect\ref{ana_RmRS2b} as seen by a viewer looking directly down the $z$ axis onto the plane of the mean Sun, the planet's equatorial plane --- shown here by the connected dots.} 
\pptfig{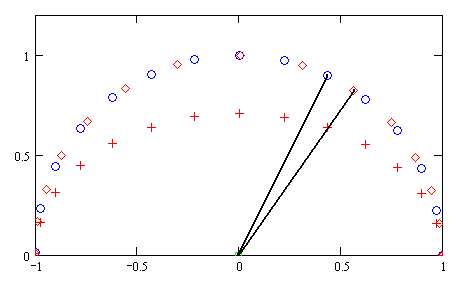}{The mean Sun (o) moves at a steady rate around the celestial equator. The tilted orbit's projection onto the plane of the mean Sun (Earth's equatorial plane) is shown by the +'s.  The $\diamond$'s are the projection of the +'s onto the celestial equator, and they move back and forth relative to the mean Sun, i.e., relative to the regular time of a clock. (The tilt in this figure is $45\,\dg$.)} Figures~\ref{ana_tiltz} and \ref{ana_tilt5} show how the tilt causes the real Sun to vary in celestial longitude relative to the mean Sun.  It is worth examining these figures closely both to understand the geometric origins of the effect and because they clarify how to extract a quantitative account of the effect from the two vectors $\vec{r}_{\rm S}(t)$ and $\vec{r}_{\rm m}(t)$. 
 
This difference in the longitudes of the two meridians can be represented by a length of arc on the equatorial path of the mean Sun.  To get this arc length, project $\vec{r}_{\rm S}(t)$ onto the equatorial plane, and then extend the projection to intersect the circle of the mean Sun. Figures~\ref{ana_tiltz} and \ref{ana_tilt5} illustrate the construction of the arc length, and Fig.~\ref{ana_tilt5} shows how the movement of the Sun causes this arc length to vary as the real Sun moves back and forth relative to the meridian of the mean Sun.   Combined with the seasonal change in declination this variation in arc length produces an analemma that is a figure eight. 

Why a figure eight?  The geometry of Figs.~\ref{ana_tiltz} and \ref{ana_tilt5} answers the question.  These figures show the relation between the mean Sun and the actual Sun over a succession of days.  On the equinox an observer sees them both at the same meridian at the same time.  But Fig.~\ref{ana_tiltz} shows that this alignment is soon lost.  The figure is a view along the $z$ axis perpendicular to the plane of the mean Sun, i.e., the equatorial plane.  See how the tilt foreshortens the projection. This projection is shown by the +'s in Fig.~\ref{ana_tilt5}, i.e., the +'s are the projection of the tip of $\vec{r}_{\rm S}$ onto the plane of $\vec{r}_{\rm m}(t)$.  An observer will see an angular separation that corresponds to the arc between the extension (the $\diamond$'s) of the projection of $\vec{r}_{\rm S}$ onto the arc along which the mean Sun is displayed (the o's).
  
In Fig.~\ref{ana_tilt5} the time interval between successive o's, +'s, and $\diamond$'s is the same ($\sim 13$~d), yet, as the figure shows, the $\diamond$'s move back and forth relative to their corresponding o's.\footnote{The number of o's is the same as the number of +'s is the same as the number of $\diamond$'s.} To check this statement place a straight edge on the box ($\Box$) in Fig.~\ref{ana_tilt5} that marks the center of the graph, and notice that a line from the box to a + connects directly to a $\diamond$. This $\diamond$ is the projection of the + onto the equatorial circle.  If you now follow the behavior point-by-point, you can see that at the equinox the $\diamond$ lines up with its corresponding o --- the mean Sun and the actual Sun cross the meridian at the same time.  But then, at first, the Sun lags behind the mean Sun. These lags accumulate until the mean Sun is half way to the solstice ($45\,\dg$ on the diagram); after that point the $\diamond$ moves faster than its corresponding o, and catches up to it so they both reach $90\,\dg$ together.  The $\diamond$ passes its o, but after $135\,\dg$ the $\diamond$ slows down and they both reach $180\,\dg$ together.  
 
 The cycle repeats over the second half of the year, so the back-and-forth motion occurs twice during the time that the up-down motion --- the change in declination --- goes through its cycle corresponding to one period of the orbit.  Back-and-forth twice for up-and-down once makes a figure eight. 

The preceding argument and conclusions are implicit in Eq.~\ref{eq:A}.  The declination angle $\delta$ is the angle between $\vec{\cal{A}}$ and its projection onto the equatorial plane.  If $\khat$ is a unit vector in the direction of the Earth's axis of rotation, then
\begin{eqnarray}
\cos \left(\frac{\pi}{2} - \delta\right) & = & \khat \cdot \vec{r}_{\rm S}  \nonumber \\
\sin \delta & = & \khat \cdot \vec{r}_{\rm S},  \label{eq:decl2}
\end{eqnarray} 
where  $\khat$ and $\vec{r}_{\rm S}$ are unit vectors.  

To extract a numerical result from Eq.~\ref{eq:decl2}, you need representations of $\vec{r}_{\rm S}$ and $\vec{r}_{\rm m}$ in the same Cartesian coordinate system.  For a frame with $\khat$ parallel to Earth's rotation axis and $\ihat$ and $\jhat$ in the equatorial plane, $\vec{r}_{\rm  m}$ is as given in Eq.~\ref{eq:rm}. For $\vec{r}_{\rm S}$, however, Eq.~\ref{eq:S} is correct only for the frame designated by primes in Fig.~\ref{rotated_axes4}, in which $\vec{r}_{\rm S}$ lies in the $\ihat'$--$\jhat'$ plane which is perpendicular to the $z'$ axis, and, therefore, perpendicular to  $\khat'$.    
\pptfig{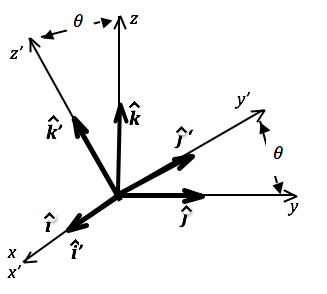}{The connection between the unit vectors $\ihat$, $\jhat$, $\khat$,  and the primed unit vectors of a frame rotated an angle $\theta$ about their shared $x$-axis} 

The angle $\theta$ between $\khat$ and $\khat'$ is the planet's tilt; $\theta= 23.44$\,\dg for Earth.  For the case where one plane is tilted by rotating it an angle $\theta$ around the $x$-axis, the connection between the primed and unprimed unit vectors is, as you can see from Fig.~\ref{rotated_axes4},
\begin{eqnarray}
\ihat' & = & \ihat \nonumber \\
\jhat' & = & \cos\theta \jhat + \sin\theta \khat \nonumber \\
\khat' & = & -\sin \theta \jhat + \cos\theta \khat.  \label{eq:rot}
\end{eqnarray}
 Using Eqs.~\ref{eq:rot} to express $\vec{r}_{\rm S}(t) =  \cos \omega t \ihat' +  \sin \omega t \jhat'$ in terms of the unprimed reference frame,\footnote{You can write down Eq.~\ref{eq:tilted_orbit} directly without going through the transformation, but it is good to use this transformation here because it will be used two more times in constructing Earth's analemma.} you will get
\begin{eqnarray}
\vec{r}_{\rm S}(t)  &=  &  \cos \omega t \ihat \nonumber \\
  & & \mbox{\ \ } + \sin\omega t \cos\theta \jhat \nonumber \\
 & & \mbox{\ \ \ \ \ \ \ }+  \sin\omega t \sin\theta \khat. \label{eq:tilted_orbit}
\end{eqnarray}
 
 Now  you can evaluate Eq.~\ref{eq:decl2} and find the declination as a function of time:
 \begin{eqnarray}
 \sin \delta & = & \sin \omega t \sin \theta \nonumber \\
   & = & \sin n\omega_0 \sin \theta \nonumber \\
 \delta & = & \sin^{-1}\left( \sin n\omega_0 \sin \theta \right).  \label{eq:decl3}
 \end{eqnarray}
 Notice that, as expected, over the course of a year, i.e., from $n=1$ to $n=365$, $\delta$ varies from $+\theta$ to $-\theta$ and back again. 
 
What about the variation of the difference along the equator? To find this follow the steps used to obtain Fig.~\ref{ana_tilt5}.  First, project $\vec{r}_S(t)$ onto the plane of the mean Sun. The projection is the $\ihat$ and $\jhat$ components of Eq.~\ref{eq:tilted_orbit}, so you just omit the $\khat$ component and get the projected vector $\vec{r}_{Sxy}$ where 
\begin{eqnarray}
\vec{r}_{\rm Sxy}(t)  &=  &  \cos \omega t \ihat \nonumber \\
  & & \mbox{\ \ } + \sin\omega t \cos\theta \jhat  \label{eq:rSxy}
  \end{eqnarray}
 
 This vector locates the +'s in Fig.~\ref{ana_tilt5}.  But you want to compare to the o's not the +'s but the $\diamond$'s, i.e., the points located by the extension of $\vec{r}_{Sxy}$, to the unit circle defined by $\vec{r}_{\rm m}$. To make this extension,  scale $\vec{r}_{\rm Sxy}$ to make it be the same length as $\vec{r}_{\rm m}$. Because $\vec{r}_{\rm m}$ is of unit length, you scale $\vec{r}_{\rm Sxy}$ by dividing by its magnitude $|\vec{r}_{\rm Sxy}| = \sqrt{\vec{r}_{\rm Sxy}\cdot \vec{r}_{\rm Sxy}}$, which from Eq.~\ref{eq:rSxy} is
\begin{equation}
 |\vec{r}_{{\rm S}xy}| =  \sqrt{\cos^2\omega t+\sin^2\omega t \cos^2\theta}, \label{eq:CircNorm}
 \end{equation} so that  \begin{equation}
\hat{r}_{\rm S xy} = \frac{\cos\omega t \ihat + \sin\omega t \cos\theta \jhat}{ \sqrt{\cos^2\omega t+\sin^2\omega t \cos^2\theta}}. \label{eq:unit_tilt}
 \end{equation}
 
 Equation~\ref{eq:unit_tilt} describes a unit vector $\hat{r}_{\rm Sxy}$ that is the same length as the unit vector $\vec{r}_{\rm m}$.  In Fig.~\ref{ana_tilt5} the tip of $\hat{r}_{\rm Sxy}$ constructed in this way lies on the arc defined by $\vec{r}_{\rm m}(t)$. 

The arc length on the unit circle equals (in radians) the angle, call it $\alpha$, between $\vec{r}_{\rm m}$ and $\hat{r}_{\rm Sxy}$.  You can find $\cos \alpha$ from the scalar product of the two unit vectors
\begin{eqnarray}
\cos \alpha & = &\vec{r}_m \cdot \frac{\vec{r}_{\rm S xy}}{|\vec{r}_{\rm Sxy}|}  \nonumber \\
&= &  \frac{ \cos^2\omega t + \sin^2\omega t \cos\theta }{\sqrt{\cos^2\omega t+\sin^2\omega t \cos^2\theta}}.
\end{eqnarray}

Alternatively, you can use the vector cross product to find $\sin\alpha$. 
\begin{eqnarray}
 \sin\alpha  & = &\khat \cdot (\vec{r}_m \times \vec{r}_{\rm S xy})  \nonumber \\
&=& \frac{\sin\omega t\cos\omega t (\cos\theta-1)}{\sqrt{\cos^2\omega t + \sin^2\omega t \cos^2\theta}} \nonumber \\
& = & \frac{- \sin(2\omega t) \sin^2(\frac{\theta}{2})}{\sqrt{\cos^2\omega t + \sin^2\omega t \cos^2\theta}}. \label{eq:eot}
 \end{eqnarray}  
 
 Using the cross product has some benefits.  One is that when $\vec{r}_{\rm m}$ and $\vec{r}_{\rm S}$ are nearly equal, $\alpha$ is small enough so that $\sin \alpha \approx \alpha$.  Another is that application of the double-angle trig identity makes it evident that the abscissa varies with twice the frequency of the ordinate.
  
Equations~\ref{eq:decl3} and \ref{eq:eot} are parametric equations of the analemma,  \pptfig{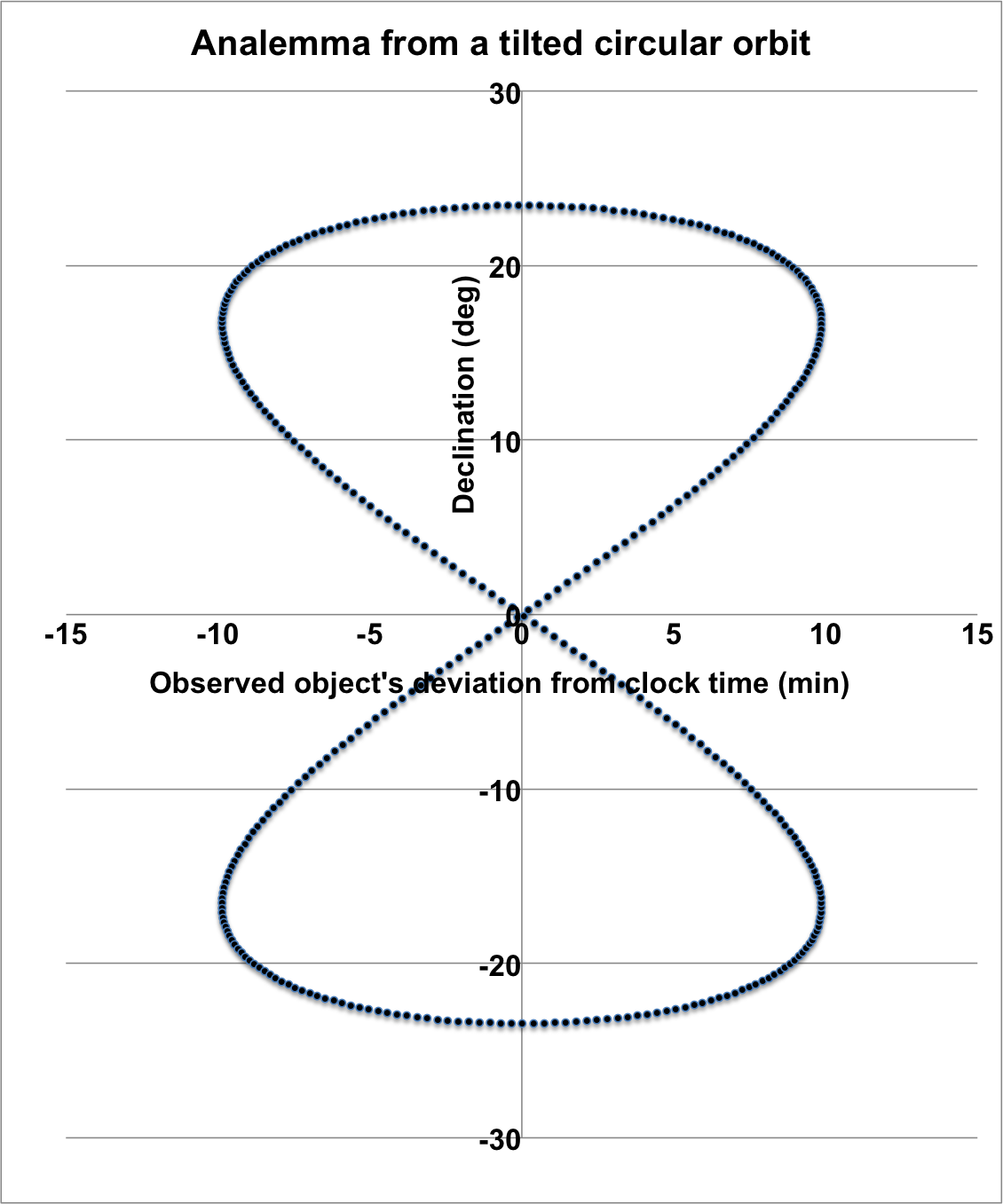}{Earth's analemma if Earth's orbit were a circle. Each dot is 1~d.} and they produce the analemma curve shown in
  Fig.~\ref{ana_circles}. This plot of declination $\delta$\ vs.\ $\alpha$, the difference between the longitude of a Sun with a circular path and the longitude of the mean Sun, was made using Excel.  As foreseen, it is a symmetric figure eight.   As is customary, the units of the abscissa are in minutes of time rather than in radians or degrees.\footnote{The conversion of  $\alpha$ from radians to minutes of time is
$\mbox{ 1 rad} \frac{180 \mbox{ deg}}{\pi \mbox{ rad}} \frac{60 \mbox{ min/h}}{15 \mbox{ deg/h}}  = 229.2 \mbox{ min}$.} 
 
 \section{Effect of orbital ellipticity on the analemma \label{ellipse}}
Because Earth's orbit is not a circle but an ellipse with the Sun at one focus, the actual analemma (shown in Figs.~\ref{ana_CO_2004}, \ref{ana_overlapped},  and \ref{analemma_dated}) differs noticeably from the circular orbit's analemma shown in Fig.~\ref{ana_circles}.  The effects of the ellipticity can be included by replacing $\vec{r}_{\rm S}(t)$ for a tilted circular orbit (Eq.~\ref{eq:tilted_orbit}) with $\vec{r}_{\rm S}(t)$ for the actual elliptic orbit.  

A good way to get the $\vec{r}_{\rm S}$ that corresponds to the correct shape of Earth's orbit is to numerically integrate the equations of motion of Earth around the Sun to obtain $\vec{r}_{\rm E}(t)$, a vector that traces out Earth's orbit, and then use the fact that $\vec{r}_{\rm S}(t) = -\vec{r}_{\rm E}(t)$. 

\pptfig{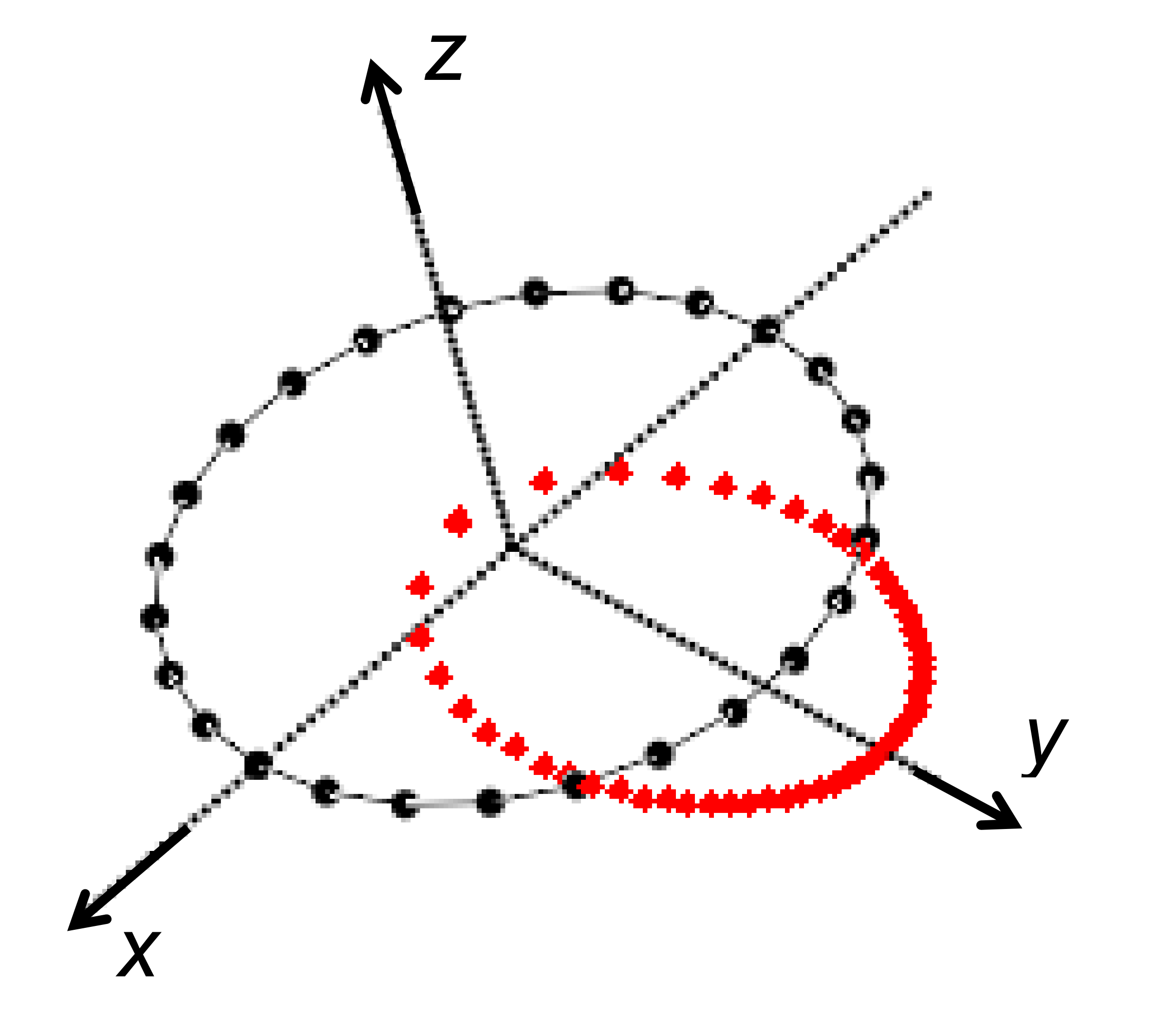}{Step 1: With the orbit (diamonds) in the plane of the mean Sun (connected dots) set the coordinate axes to have their origin at the focus of the ellipse and the $y$-axis along the ellipse's major axis $2a$.  Perihelion is at $y=-a(1-\epsilon)$.} 
\pptfig{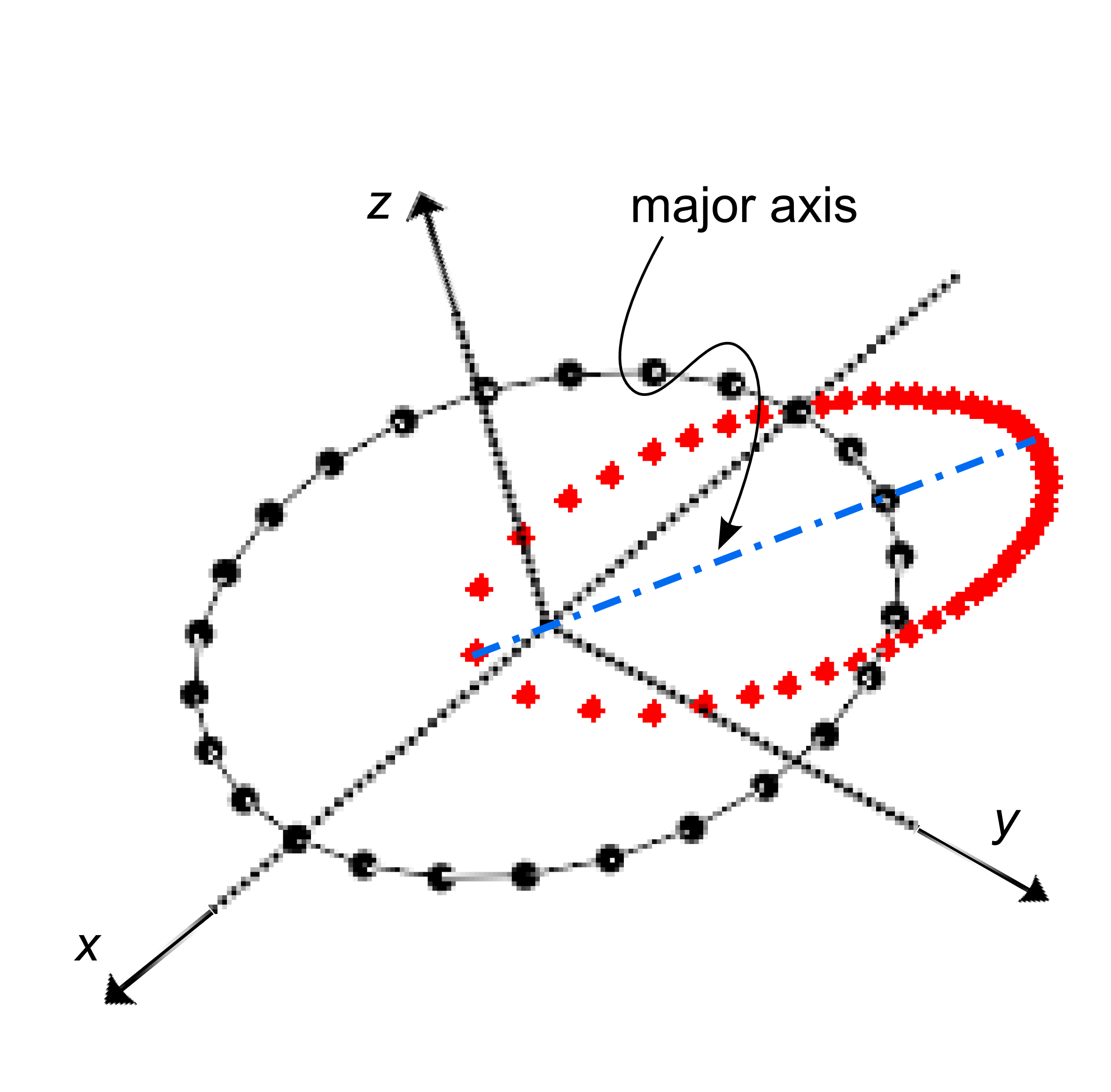}{Step 2: Rotate the axes to align the $x$-axis with the orbit's line of nodes.  In this picture, the axes have been rotated around the $z$-axis by $-60\,\dg$ so the major axis of the ellipse is now at $60\,\dg$ to the $y$-axis. The line of nodes is the segment of the $x$-axis crossing the ellipse.}
\pptfig{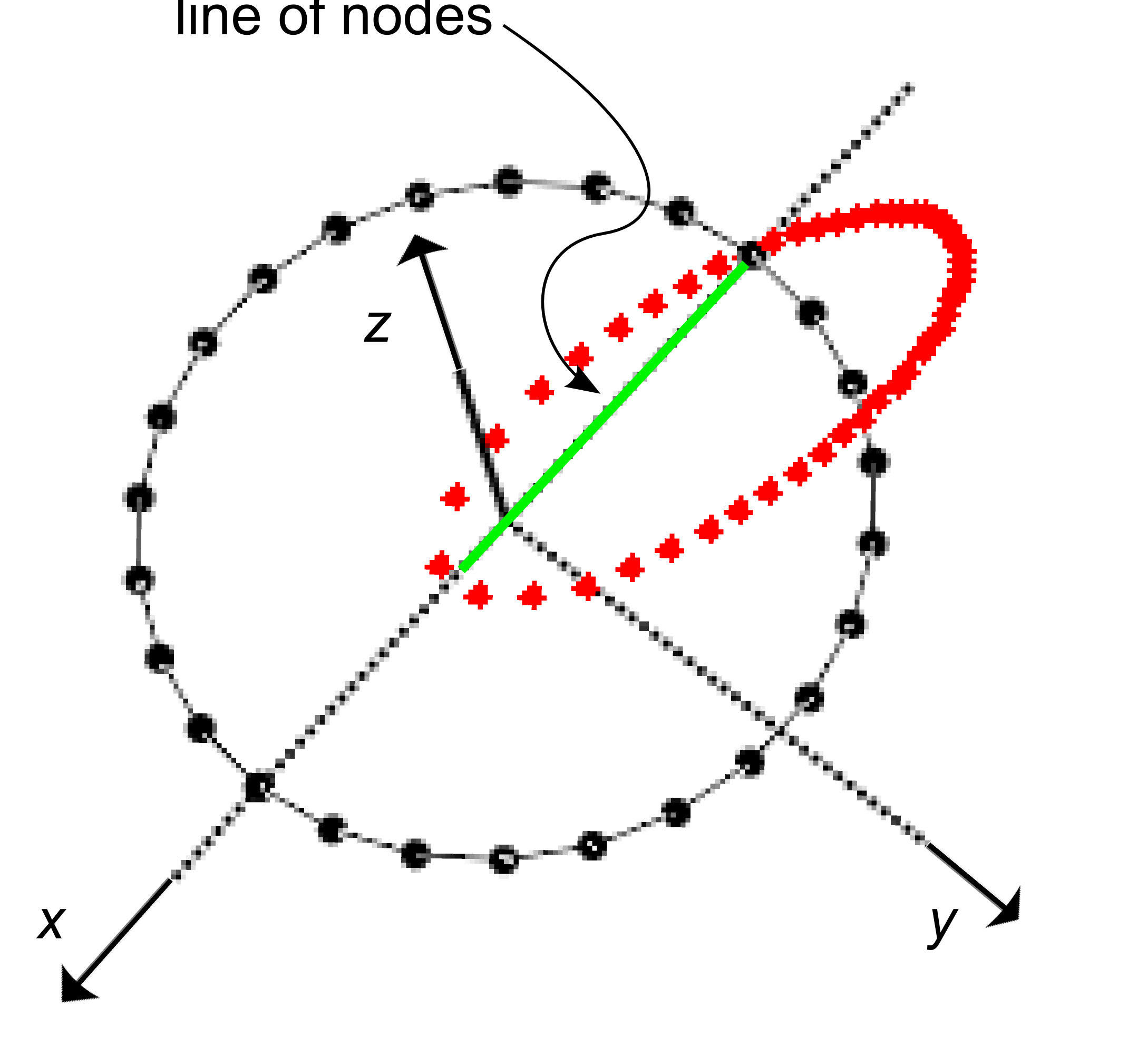}{Step 3: Rotate the axes around the $x$-axis so that the orbital plane is tilted $23.44\,\dg$ around the line of nodes (solid line).  To make the tilt more apparent it has been set to $40\,\dg$ in this picture.}

To do this integration you need the right initial conditions.  A good way to find them proceeds in three steps.  Step one,  find the initial conditions in a coordinate frame aligned with the axes of the ellipse. This orientation of the axes makes determining the initial conditions particularly simple.  

Step two, transform the initial conditions determined in step one into the frame of reference that lines up its $x$-axis with the line of nodes --- the line of intersection of the orbital plane with the plane of the circular orbit of the mean Sun (see Fig.~\ref{ana_RmRS2b}).  For Earth, this corresponds to a rotation of the orbital plane $12.8\,\dg$ about the $z$-axis.  

Step three, transform the initial conditions to correspond to the orbital plane tilted around its line of nodes. 

 Numerical integration with the initial conditions from step three gives a vector $\vec{r}_{\rm E}$ that accurately represents Earth's orbit.  The resulting $\vec{r}_{\rm S}$, which equals $-\vec{r}_{\rm E}$, is no longer the unit vector of Eq.~\ref{eq:S} sweeping out a circle.  Now it sweeps out an ellipse with eccentricity $\epsilon = 0.0167$ tilted around the line of nodes at an angle of $\theta = 23.44\,\dg$ relative to the plane of the mean Sun, speeding up at perihelion and slowing down at aphelion.  This vector $\vec{r}_{\rm S}$, the mean Sun vector  $\vec{r}_{\rm m}$, and slightly modified versions of Eqs.~\ref{eq:decl2} and \ref{eq:eot} yield a quantitatively accurate representation of Earth's analemma.  

 \pptfig{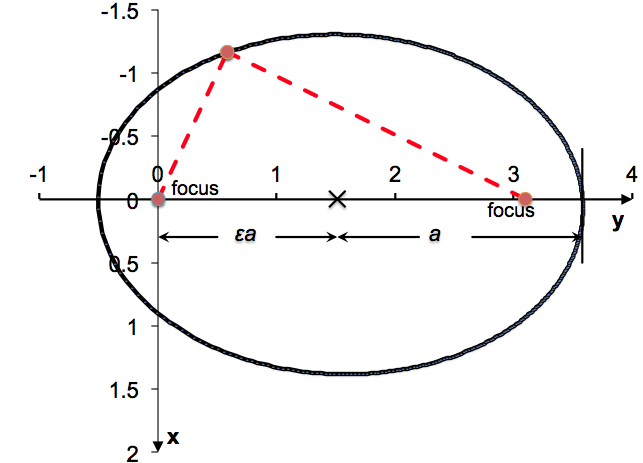}{The sum of the distances from any point on the curve of an ellipse to the two points labeled ``focus'' is a constant equal to the length of the major axis $2a$. The eccentricity $\epsilon$ is the distance from the center of the ellipse to either focus, measured in units of the semi-major axis.} 
Construction of this more elaborate $\vec{r}_{\rm S}$ vector requires some basic properties of ellipses summarized in Fig.~\ref{ellipse-parms2}. The two points labeled ``focus'' are key.  The curve is defined by the property that the distance from one focus to   any point on the curve plus the distance from that point to the other focus is constant.  (Use a ruler to check that this statement is true for Fig.~\ref{ellipse-parms2}.)  Another important feature is the eccentricity $\epsilon$.  It measures the distance of the focus from the center of the ellipse in units of the semi-major axis $a$.  (Use a ruler to confirm that the eccentricity of the ellipse in Fig.~\ref{ellipse-parms2} is $\epsilon = .75$.) This value was chosen to be much larger than the 0.0167 eccentricity of Earth's orbit to make the ellipticity more apparent in the figure. 
 
 It follows from the general properties of an ellipse that
 \begin{eqnarray}
 \mbox{ semi-minor axis: }b & =&  a \sqrt{ (1- \epsilon^2)} \;\;\; \;\;\;\; \label{eq:semi-minor} \\
 \mbox{ area of ellipse: }A & = & \pi a b   \label{eq:ellipse_area}.
 \end{eqnarray}   
 
 Newton's law of universal gravitation states that the force $\vec{F}_{21}$ exerted on a spherical mass $m_2$ by a spherical mass $M$ is
\begin{equation}
\vec{F}_{21} = -\frac{GMm_2}{|\vec{r}_2-\vec{r}_1|^2} \frac{(\vec{r}_2-\vec{r}_1)}{|\vec{r}_2 - \vec{r}_1|}  \label{eq:UniGrav}
\end{equation}
where $\vec{r}_1$ and $\vec{r}_2$ are the position vectors of the two masses, and $G$ is the constant of universal gravitation. This equation is the inverse square law of gravitational attraction multiplied by the unit vector $(\vec{r}_2-\vec{r}_1)/(|\vec{r}_2-\vec{r}_1|)$, which points along the line between the centers of the two masses. The unit vector asserts that this is a central force, and the minus sign at the beginning of the equation points the unit vector from $m_2$ to $M$ to reflect the fact that $M$ is attracting $m_2$. 

For $ M >> m_2$ and with $\vec{r} =\vec{r}_2-\vec{r}_1$, Newton's laws of motion and Eq.~\ref{eq:UniGrav} lead to a differential equation for the orbit:
\begin{equation}
 \vec{\ddot{r}} = \frac{-GM}{|\vec{r}|^2} \frac{\vec{r}}{|\vec{r}|}.  \label{eq:orbit}
\end{equation}
The orbiting mass $m_2$ has energy $E$ where
\begin{equation}
\frac{E}{m_2} = \frac{1}{2} \vec{v}\cdot \vec{v} -  \frac{GM}{|\vec{r}|}, \label{eq:Energy}
\end{equation} and angular momentum $\vec{L}$ such that
\begin{equation}
\frac{\vec{L}}{m_2} = \vec{r} \times  \vec{v} = r^2 \vec{\omega},  \label{eq:AngMom}
\end{equation} where $\vec{\omega}$ is the angular velocity of the planet around the Sun.  $E$ and $L$ are conserved (constant) quantities.

For initial values of  position and velocity such that the energy $E<0$, the solutions to the differential equation are orbits that are ellipses with semi-major axis $a$ and the Sun at one focus.  The following equations connect the geometry of the elliptical orbit to the underlying physics:
\begin{eqnarray}
\epsilon  &=& \left(1 + \frac{2EL^2}{G^2M^2}\right)^{\frac{1}{2}} \label{eq:epsilonEL} \\
a & = &\left |\frac{GM}{2E}\right|  \label{eq:semimajorE} \\
T^2 & = & \frac{4\pi^2}{GM} a^3  \label{eq:K3}
\end{eqnarray}
where $T$ is the period of the planet's orbit.\cite{*[{}] [{ (see pp. 111-117).}] Sym:53}   Notice that with $a$ measured in AU and $T$ in years,  Eq.~\ref{eq:K3} becomes $T^2 = 1 = 4\pi^2/GM$ for Earth, so that in these units $GM = 4 \pi^2$ when $M$ is the mass of the Sun. 

These equations are interesting for what they say about planetary motion. They are essential for checking the validity of orbits produced by numerical integrations. They are also useful for obtaining initial conditions for orbits with parameters different from those of Earth's orbit.

\begin{table}
\caption{Some Parameters of Earth's Orbit \label{tb:EarthOrbitParms}}
\begin{tabular}{|c|p{1.2in}|rl|} 
\hline
  $\epsilon$ & Eccentricity & 0.0167 & \\ \hline
  $a$ & Semi-major axis & 1 & AU \\ \hline
  $T$ &Period & 1 & y  \\ \hline
  $E$ & Energy  per solar mass & -19.739 & AU$^2$/ y$^2$ \\ \hline
  $L$ &Angular momentum per solar mass & 6.2823 & AU$^2$/y \\ \hline
  $\phi$ & Angle from line of nodes to perihelion  & -78.2 &deg \\ 
\hline

\end{tabular}
\end{table}

To find initial conditions that correspond to Earth's orbit, that is, to an orbit that has $\epsilon = .0167$ and the other parameters given in Table~\ref{tb:EarthOrbitParms}, start with step one of the procedure outlined above.  

In a Cartesian coordinate frame on the ellipse with the origin at one focus, the $x$-axis coincident with the ellipse's minor axis, and the $y$-axis with its major axis, Earth's position vector at perihelion is $\vec{r}_{\rm E} = 0 \ihat - a(1-\epsilon) \jhat$ and its velocity is $\vec{v}_{\rm E} = v_{\rm p} \ihat + 0 \jhat$. 

The quantity $v_{\rm p}$ is the magnitude of Earth's velocity at perihelion. You can determine $v_{\rm p}$ using  Kepler's law of equal areas in equal times (conservation of angular momentum $\vec{L}$).  

A vector $\vec{r}$ from the Sun to a point moving with velocity $\vec{v}$ sweeps out area at a rate
\begin{equation}
\frac{d\vec{A}}{dt} = \frac{1}{2} \vec{r} \times \vec{v}.  \label{eq:SweptArea}
\end{equation}  This is essentially the formula for the area of a triangle --- 1/2 the base times the altitude where the base is the arc length swept out in a time $dt$.  Kepler and Newton tell us that for a central force, the quantity $d\vec{A}/dt$ is constant in both magnitude and direction.  It follows that for Earth the value for that constant must be the same as the total area swept out in a year, i.e., Eq.~\ref{eq:ellipse_area}, divided by the time of one year:
\begin{eqnarray}
\left|\frac{d\vec{A}}{dt}\right| & = & \frac{\pi a b}{T}
 =  \frac{\pi a^2\sqrt{1-\epsilon^2}}{T}  \nonumber \\
 &=& 3.141 \mbox{ AU$^2$/y}.
\end{eqnarray}
But, as  Eq.~\ref{eq:AngMom}  states, $\vec{r} \times \vec{v}$ is also the angular momentum $\vec{L}$ of a unit mass. (For any given initial conditions, the orbit is independent of the mass of the orbiting object, so set $m_2 =1$ to simplify the discussion.) 
\begin{eqnarray}
\frac{L}{m_2} \equiv L & = & 2 \frac{dA}{dt} \nonumber \\ 
L_{\rm Earth} & = & 6.282 \mbox{ AU$^2$/y}
\end{eqnarray} And, because at perihelion $\vec{r}_{\rm E}$ and $\vec{v}_{\rm E}$ are exactly perpendicular,
\begin{eqnarray}
L_{\rm Earth} & = & a (1-\epsilon) |\vec{v}_{\rm E}| \nonumber \\
 & =&  (1 -.0167)\; {v}_{\rm p} \nonumber \\
 & = & .9833\;  {v}_{\rm p} =  6.282,
\end{eqnarray} 
and the magnitude of the velocity at perihelion is
\begin{equation}
v_{\rm p}= 6.389 \mbox{ AU/y}.
\end{equation}

In the chosen reference frame the position of Earth at perihelion is $\vec{r}_{\rm E} = -.9833 \jhat$~AU, and its velocity is $\vec{v}_{\rm E} = 6.389 \ihat$~AU/y.  It would be equally convenient to determine the initial conditions at aphelion.  

With these initial conditions an Euler integration of the differential equation with an initial half-step gives a good result.  This is the method that Feynman used in his lectures.\cite{*[{}] [{ pp. 9-6 -- 9-9}] Feyn1:63}  In the units used here a convenient time step for the integration is $\Delta t = 1/365.25 = .002738$~y.  This is the duration of 1~d (one day) so 366 steps of integration produce a complete orbit, and with this time step you can use the representation of $\vec{r}_{\rm m}$ given in Eq.~\ref{eq:rm}.  The integration can be done with any mathematical software package. A spreadsheet works well, and the calculation used here is available with the online version of this paper.

\begin{table}[htbp] \caption{Initial values of $\vec{r}_{\rm E}$ in AU and $\vec{v}_{\rm p}$ in AU/y for calculating Earth's orbit in three different orientations \label{tb:3InitVals}}
 \begin{tabular}{|c|c|c|c|} \hline  
\multicolumn{1}{|p{1.in}|}{\vspace{-2cm}Orientation of ellipse relative to coordinate axes} &  {\begin{sideways}
      Perihelion on $-y$ axis \ \ \ 
      \end{sideways}
      }
 & {\begin{sideways}Orbit plane rotated $-12.8\,\dg$ \end{sideways}} 
 & {\begin{sideways}Orbit plane tilted $23.44\,\dg$ \end{sideways}} \\
  \hline \hline
  $x_0$ & 0 & 0.2184 & 0.2184 \\ \hline
  $y_0$ & -0.9833 & -0.9588 & -0.8796 \\ \hline
  $z_0$ & 0 & 0 & 0.3814 \\ \hline \hline
  $v_{x01/2} $ & 6.3890 & 6.2171 & 6.2171 \\ \hline
  $v_{y01/2}$ & 0.05589 & 1.4733 & 1.3517 \\ \hline
  $v_{z01/2}$ & 0 & 0 & -0.5643 \\ \hline \hline 
 \end{tabular} 
  \end{table}

Integration with these initial conditions starts from perihelion (January 3) and gives the day-by-day position of Earth on its orbit with the specified eccentricity. However, this orbit needs to be correctly oriented relative to the idealized orbit of the mean Sun.    

Do this with steps two and three.  Step two: make the line of nodes coincide with the $x$-axis by rotating the coordinate frame about its $z$-axis until Earth crosses the $x$-axis on March 20, the vernal equinox of the Northern Hemisphere.  

To find the angle of rotation use orbit data produced by numerical integration with a tilt angle but no rotation of the orbit plane. These data show Earth crossing the $x$-axis 13~d later than the actual March 20 equinox. This result means you must rotate the reference frame clockwise by $-13/365.25\times 360 = -12.8\,\dg$ relative to the orbit to place the $x$-axis so that Earth crosses it on March 20.\footnote{You can also use the fact that perihelion for Earth does not occur on  the day of the solstice, December 21, as would be the case if you did not rotate the axes.  Perihelion occurs 13 days later on January 3, so the rotation of the axes that puts perihelion at the proper date is (as it should be) the same as the rotation that places the March equinox on the $x$-axis. Another alternative is to look up the celestial coordinates of perihelion; see, for example, \protect\url{<http://nssdc.gsfc.nasa.gov/planetary/factsheet/earthfact.html>} }

To rotate the coordinate frame, use the following version of  Eq.~\ref{eq:rot} modified to take into account that this rotation is about the $z$-axis:\footnote{You can generate Eqs.~\protect\ref{eq:rotz} from Eq.~\protect\ref{eq:rot} by permuting $x$, $y$, and $z$ (and $\ihat$, $\jhat$, and $\khat$.)}
\begin{eqnarray}
x_0 &=& x_0' \cos\alpha + y_0' \sin\alpha \nonumber \\
x_0 & = & 0 + (-0.9833 \times -0.2221) = 0.2184  \nonumber \\
y_0 &=& -x_0' \sin\alpha + y_0' \cos\alpha \nonumber \\  
y_0 & = & 0 + (-0.9833 \times 0.9750) = -0.9588.  \nonumber  \label{eq:rotz}
\end{eqnarray}
These equations produce the second set of initial conditions in Table~\ref{tb:3InitVals}.

Step three: tilt the orbital plane.  Do this by applying Eq.~\ref{eq:rot} to the set of orbital coordinates labeled ``rotated'' in Table~\ref{tb:3InitVals}; you will get the initial conditions labeled ``tilted'' in Table~\ref{tb:3InitVals}. Integration with these values gives Earth's orbit tilted relative to the equatorial plane.  

\begin{table}\caption{ Selected values of $\vec{r}_{\rm E}$ from the numerical integration of Earth's orbit.} \label{outputs}
\begin{tabular}{|c|r|r|r|r|} \hline
Date &	$x_n$& $y_n$ & $z_n$ & $r_n$ \\ \hline \hline
3-Jan & 0.2184 &	-0.8796 &	0.3812 &  	0.9833 \\ \hline
1-Feb &  0.6638 & -0.6682 &	0.2897 & 	0.9854 \\ \hline 
20-Mar & 0.9959 &  -0.00647 & 0.00280 &   	0.9960 \\ \hline
21-Mar & 0.9962 & 0.00940 &  -0.00407 &  0.9963 \\ \hline
20-Jun  &  0.02009 &	0.9323 & -0.4042 &  	1.0163 \\ \hline
21-Jun & 0.00317 & 0.9325 & -0.4043 & 	1.0164 \\ \hline
22-Jun &  -0.0138 & 	0.9325 & -0.4043 &	1.0165 \\ \hline
5-Jul &  -0.2317 & 0.9084 & -0.3953 & 1.01685 \\ \hline
22-Sep  & -1.00366 & 0.01344 & -0.00583 & 1.00377 \\ \hline
23-Sep  & -1.00349 & -0.00229 & 0.00099 &  1.00349 \\ \hline 
\end{tabular}
\end{table}

The result of the numerical integration is a table of values of $x_n$, $y_n$, and $z_n$, the Cartesian components of $\vec{r}_{\rm E}(n \omega_0)$, the vector that locates Earth relative to the Sun on day $n$ since perihelion:
\begin{equation}
\vec{r}_{\rm E}(n \omega_0)  =  x_n \ihat +y_n  \jhat + y_n \khat 
 \label{eq:tilted_ellipse}
\end{equation}  

Some results from the numerical integration are given in Table~\ref{outputs}. These should make you confident that the calculation is accurate. Between March 20 and 21  $y_n$ and $z_n$ pass through zero as they should at the vernal equinox of the Northern Hemisphere.  And between June 21 and June 22, when the summer solstice occurs, $x_n$ changes sign as it should.  The entry dated July 5 was selected because on that date the distance $r_n$ has its largest value; in other words the calculation gives the correct date for Earth's aphelion. 

 But notice that the aphelion distance is slightly larger than $a (1+\epsilon)$; this discrepancy reflects the limitations of the method of numerical integration and the step size used here. It's a warning that integrations over more than a year may need a smaller step size or a more accurate method of integration.

To construct the analemma, you need $\vec{r}_{\rm S}$ the vector that points from Earth to Sun. It is
\begin{eqnarray}
\vec{r}_{\rm S}(n \omega_0) & = & -\vec{r}_{\rm E}(n \omega_0) \nonumber \\ 
&=&  -x_n \ihat -y_n  \jhat - y_n \khat  \label{eq:rSellipse}
\end{eqnarray}

To extract $\alpha$ and $\delta$, modify Eqs.~\ref{eq:CircNorm}, \ref{eq:unit_tilt}, and \ref{eq:eot} to be Eqs.~\ref{eq:EllipseNorm}, \ref{eq:unit_tilt_ellipse}, and \ref{eq:eot_ellipse} and use them to get $\alpha$ and $\delta$ as functions of time, i.e., as functions of $n$  the day number. 

\begin{equation}
 |\vec{r}_{{\rm S}xy}(n\omega_0)| = \sqrt{x_n^2+y_n^2}, \label{eq:EllipseNorm}
 \end{equation} so that  \begin{equation}
\vec{r}_{\rm S xy}(n\omega_0) = \frac{x_n \ihat + y_n  \jhat}{\sqrt{x_n^2+y_n^2 }}. \label{eq:unit_tilt_ellipse}
 \end{equation}

Then
\begin{equation}
 \sin\alpha  = \khat \cdot \vec{r}_m \times \vec{r}_{\rm S xy} 
\end{equation}
so that
\begin{equation}
 \alpha_n = \sin^{-1} \left(\frac{-y_n  \cos n\omega_0 + x_n \sin n\omega_0}{\sqrt{x_n^2+y_n^2 }}\right) \label{eq:eot_ellipse}
 \end{equation} 
 and
  \begin{equation}
 \delta_n = \sin^{-1} \left( \frac{z_n}{\sqrt{x_n^2+y_n^2+z_n^2}} \right). \label{eq:decl_ellipse}
 \end{equation}
\pptfig{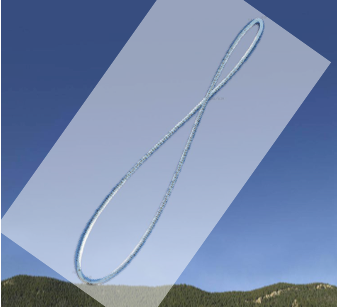}{Here the calculated analemma is displayed overlaid on the observed analemma of Fig.~\ref{ana_CO_2004}. For the best overlap the calculated analemma has been rotated through 35.6\,\dg.  This corresponds to about 2:15 pm MST in reasonable agreement with the 2:28 pm MST when the photographs of the observed analemma were taken.}\pptfig{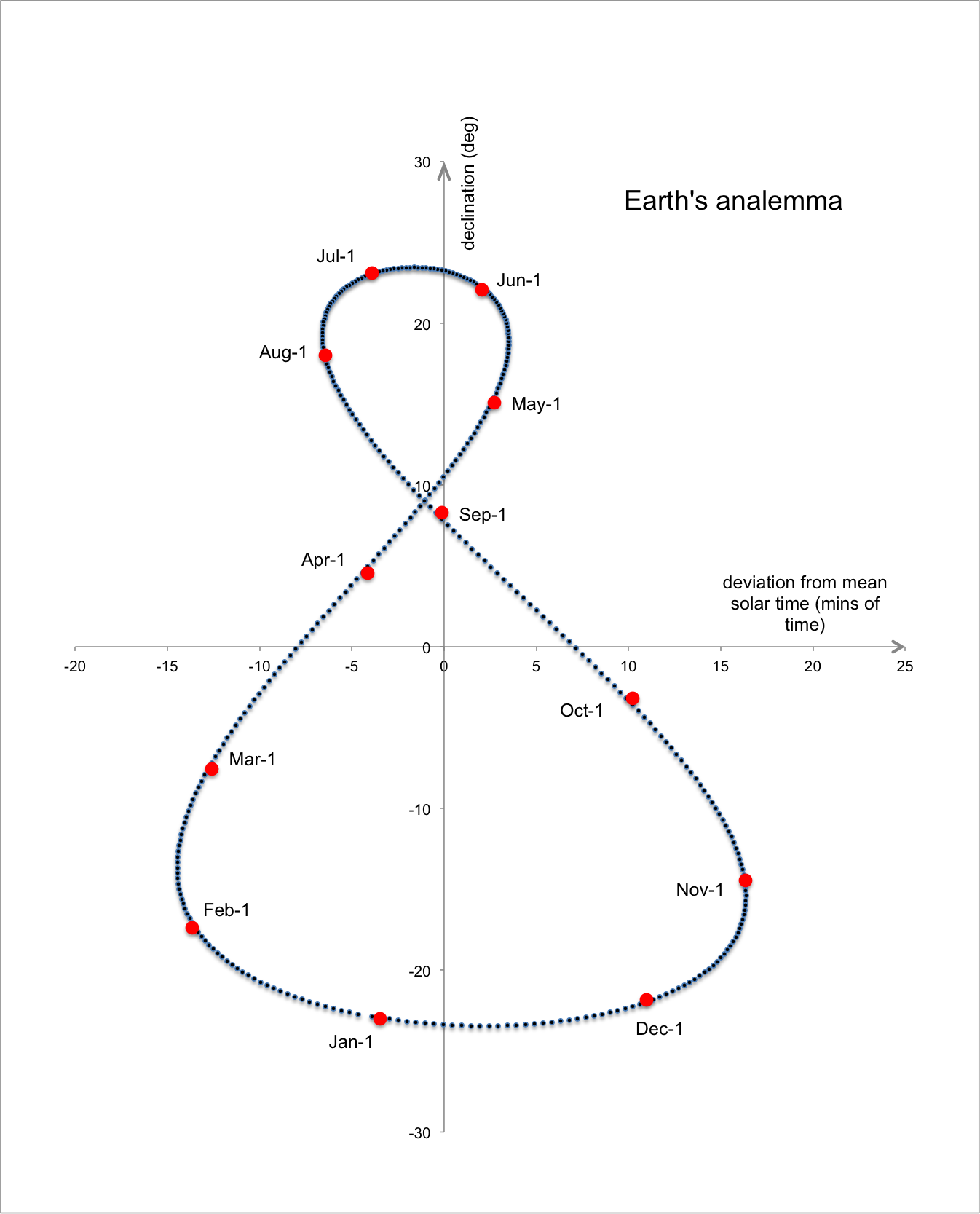}{This is the same curve as in Fig.~\ref{ana_overlapped} but oriented to be at noon and with the scale of the abscissa expanded by $\times 6$. Each dot is a different day.} 

Equation~\ref{eq:eot_ellipse} is not entirely satisfactory for extracting $\alpha$.  It is set up so that when $n=0$, $\vec{r}_{\rm S}$ points $12.8 \dg$ counterclockwise from the negative y-axis while $\vec{r}_{\rm m}$ is pointing along the x-axis.  This makes the angle between them $90-12.8=77.2 \dg$.  In principle this large angle is not important; the variation in $\alpha$ is still the correct effect; it's as though you were measuring the time at which the Sun is near its zenith with a clock in a time zone some seven or eight hours away.  However, it is tidier to set $\vec{r}_{\rm m}$ so that it corresponds to the direction of $\vec{r}_{\rm S}$.   To do this, add a phase constant $\phi = -77.2\,\dg$ so $\vec{r}_{\rm m}$ becomes 
\begin{equation} \vec{r}_{\rm m}(n) = \cos(n\omega_0 + \phi) \ihat + \sin(n\omega_0+\phi) \jhat \label{eq:rm_phase}\end{equation}
and
\begin{equation}
\alpha \approx \frac{x_n \sin (n \omega_0+\phi) - y_n \cos( n \omega_0 +\phi)}{\sqrt{x_n^2 + y_n^2 }} \label{eq:eot_phase}
\end{equation}

 The analemmas shown in Figs.~\ref{ana_overlapped} and \ref{analemma_dated} are plots of $\delta$ vs.  $ \alpha$ obtained from Eqs.~\ref{eq:decl_ellipse} and \ref{eq:eot_phase}  for $n=1$ to $n=365$ with $\phi=-77.2 \dg$. In Fig.~\ref{ana_overlapped} the calculated analemma has been laid over the observed analemma; the agreement between the two is quite satisfactory.  The analemma in Fig.~\ref{analemma_dated} is the same as in Fig.~\ref{ana_overlapped} but with the horizontal dimension enlarged by $6 \times$.  It shows the day-by-day variation of the arrival of the noontime Sun.

\section{Vectoria's Secret \label{Vectoria} }
The construction of the analemma in the preceding sections emphasizes visualization of how the analemma results from the relative motion of Earth and Sun in three dimensions. The goal was to show how the geometry connects with the physics.  

But more formal, less visual approaches are possible, and they have their merits.\cite{Chalm:1987}  Done properly they can carry you past various difficulties of establishing relative phases, of getting the algebraic signs correct, and they can provide an aid or an alternative for those who find it difficult to visualize three dimensions.  The following paragraphs make a fuller use of vector algebra to reprise the construction of the analemma.

The key question in constructing the analemma is ``What is the angle $\alpha$  between $\vec{r}_{\rm m}$ and the projection of $\vec{r}_{\rm S}$ onto the plane of $\vec{r}_{\rm m}$?" One way to find the answer is to use the fact that $\alpha$ is the dihedral angle between two planes, one defined by $\khat$ and $\vec{r}_{\rm m}$ and the other by $\khat$ and $\vec{r}_{\rm S}$.  \pptfig{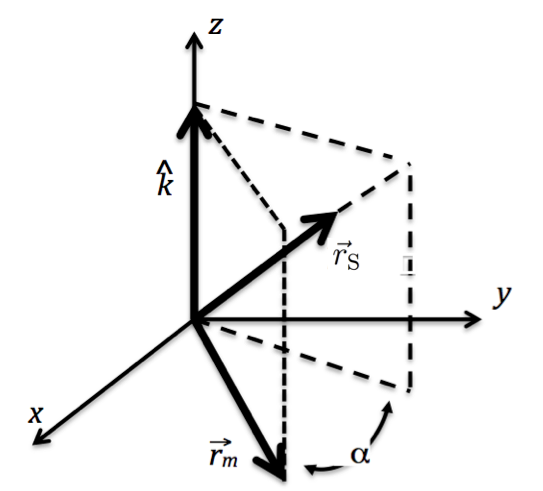}{The vectors $\khat$, $\vec{r}_{\rm m}$, and $\vec{r}_{\rm S}$ define two planes.  The variation between the positions of the Sun and the idealized mean Sun causes $\alpha$ to vary from day to day.} Figure~\ref{ana_dihedralS} shows the configuration.  

The angle $\alpha$ between the two planes is the same as the angle between unit vectors drawn perpendicular to the two planes. This fact suggests the following procedure to find $\sin \alpha$: Construct unit vectors  $\hat{n}_{\rm S}$ and $\hat{n}_{\rm m}$ normal to the two planes; take their cross product, which will give a vector in the $\khat$ direction with magnitude of $\sin \alpha$; take the dot product of $\khat$ with the cross product to find $\sin \alpha$ (see Eq.~\ref{eq:sina0}).

Use cross products to construct $\hat{n}_{\rm S}$
\begin{equation}
\hat{n}_{\rm S}  = \frac{\khat \times \vec{r}_{\rm S}}{|\khat \times \vec{r}_{\rm S}|} \label{eq:nS}
\end{equation}
and $\hat{n}_{\rm m}$
\begin{eqnarray}
\hat{n}_{\rm m} &=&\frac{ \khat \times \vec{r}_{\rm m} }{| \khat \times \vec{r}_{\rm m} |}\nonumber \\
&=&   \khat \times \vec{r}_{\rm m} \label{eq:nm}
\end{eqnarray} where Eq.~\ref{eq:nm} follows from the fact that $|\khat \times \vec{r}_{\rm m}| = 1$ because $\khat$ and $\vec{r}_{\rm m}$ are both unit magnitudes and perpendicular to each other.

This is enough information to determine $\sin \alpha$ by direct substitution into 
\begin{equation}
\sin \alpha = \khat \cdot (\hat{n}_{\rm m} \times \hat{n}_{\rm S}), \label{eq:sina0}
\end{equation} but using the vector identity $\vec{A} \cdot(\vec{B} \times \vec{C}) = (\vec{A} \times \vec{B}) \cdot \vec{C}$ and the fact that $\vec{r}_{\rm m}$ is 
\begin{equation}
\vec{r}_{\rm m} = -\khat \times \hat{n}_{\rm m}. \label{eq:rm2}
\end{equation}
 yields the more compact expression:

\begin{eqnarray}
\sin \alpha &=& \khat \cdot (\hat{n}_{\rm m} \times \hat{n}_{\rm S}) \nonumber \\
&=& (\khat \times \hat{n}_m)\cdot \hat{n}_{\rm S} \nonumber \\
&=& -\hat{n}_{\rm S} \cdot \vec{r}_{\rm m}. \label{eq:sina}
\end{eqnarray}

Substituting Eqs.~\ref{eq:rm_phase} and \ref{eq:nS}  into Eq.~\ref{eq:sina} gives
 Eq.~\ref{eq:eot_phase}.

The vector expression for the declination was already used to obtain the result given in Eq.~\ref{eq:decl_ellipse}. 

\section{Factoids and Projects \label{projects}}
The analemma occurs because solar days vary in length.  A curious consequence is that the longest solar day, about 24\;h~30\;s, occurs just after the winter solstice, the day on which in the Northern Hemisphere there are the fewest hours of daylight. Thus, ``the shortest day of the year is also the longest day of the year" --- or nearly. It is also interesting that for about three weeks after the solstice  sunrise continues to get later. Nevertheless, the amount of daylight increases in the Northern Hemisphere because sunset is getting later by an amount that more than offsets the later sunrise.  {\em The New York Times} finds these factoids fit to print.\cite{ONeil:2013}

This tutorial can be the basis for further projects. A project might be as modest as an exercise in vector manipulation.  For example, you might replace $\hat{n}_{\rm m}$ and $\vec{r}_{\rm S}$ in Eq.~\ref{eq:sina0} with their cross product definitions from Eqs.~\ref{eq:nS} and \ref{eq:nm} and use the identity $\vec{A} \times (\vec{B} \times \vec{C}) = \vec{B}\,( \vec{A}\cdot\vec{C}) - \vec{C} \, (\vec{A}\cdot \vec{B})$ to derive Eq.~\ref{eq:sina}. 

Or you could measure the declination of the Sun and observe the time at which it reaches its zenith and see if these agree with the predictions of your analemma.  You will need to understand how to make $\vec{r}_{\rm m}$ correspond to your clock time -- a worthwhile task. 

You can modify the input parameters for the numerical calculation of Earth's orbit and examine how Earth's analemma will change as the equinoxes precess.  You can also replace Earth's orbital and rotational parameters with those of Mars and find the analemma that an observer on Mars will see. By properly modifying the input parameters of the numerical integration, you can find the analemma of any planet.  

The analemma is the result of the superposition of two periodic motions at right angles to each other.  In effect it is an interference pattern, a Lissajou figure.  The rather small eccentricity of Earth has a surprisingly large effect on the analemma.  Can you invert the argument, and find the eccentricity of Earth's orbit from its observed analemma? 

A nice project would be to calculate the transits of Venus using this paper's approach. Numerically calculate the position vector $\vec{r}_{\rm V}$ of Venus relative to the Sun. Then construct a pointer from Earth to Venus $\vec{r}_{\rm EV} = \vec{r}_{\rm V}- \vec{r}_{E}$.  Look at the angular difference between $\vec{r}_{\rm S}$ ($ = - \vec{r}_{\rm E}$) and $\vec{r}_{\rm EV}$ as a function of time.  Find the times at which the angular difference is between $-1/4 \dg$ and $+1/4 \dg$, i.e., approximately within the angular diameter of the Sun.  You will need a method of numerical integration that is stable over the long interval of time between the occurrence of one pair of transits and the next pair. The Euler method will not work, but a fourth order Runge Kutta method available on most packages of mathematics software may do the job.

\section{Conclusion \label{conclusions}}
This tutorial shows how to build your own analemma using only the physics taught in calculus-level introductory physics. The result is the construction of a quantitatively accurate  analemma for Earth.  It is a concrete example of how successfully Newtonian mechanics describes the solar system.   

We tell our students that Newtonian physics is a triumph of human intellect; we describe interesting motions of planets, moons, comets, etc. and then say that Newtonian mechanics explains them. It is important for physics students to go beyond praise for the achievements of Newtonian physics. This tutorial enables them to recreate and experience some of that achievement. 

\begin{acknowledgments}
I thank Howard Georgi (Harvard University),  Ed Bertschinger and Ramachandra Dasari (MIT) for their hospitality. It has greatly facilitated my work.  I thank  Joe Amato (Colgate University) for his close reading and thoughtful suggestions for improving this paper.  The goals of this paper grew out of an exchange of ideas with Shane Larson (Utah State University), and the decision to write it was inspired by ``Astronomy's Discoveries and Physics Education," a conference supported by NSF Grant 1227800.
\end{acknowledgments}

\bibliographystyle{apsrev4-1}	
\bibliography{CHH-bib-items}
 
\end{document}